\documentclass{article}

\usepackage[dblblindworkshop, final]{neurips_2025}
\workshoptitle{AI for Music}

\usepackage[utf8]{inputenc} 
\usepackage[T1]{fontenc}    
\usepackage{hyperref}       
\usepackage{url}            
\usepackage{booktabs}       
\usepackage{amsfonts}       
\usepackage{nicefrac}       
\usepackage{microtype}      
\usepackage{xcolor}         
\usepackage{multirow}

\usepackage{natbib}
\title{Ethics Statements in AI Music Papers: The Effective and the Ineffective}

\author{
  Julia Barnett\\
  Northwestern University\\
\texttt{juliabarnett@u.northwestern.edu} \\
  \And
  Patrick O'Reilly \\
  Northwestern University \\
\texttt{patrick.oreilly2024@u.northwestern.edu} \\
  \AND
  Jason Brent Smith \\
  Northwestern University \\
\texttt{jason.smith1@northwestern.edu} \\
  \And
  Annie Chu \\
  Northwestern University\\
  \texttt{anniechu@u.northwestern.edu} \\
  \And
  Bryan Pardo\\
  Northwestern University\\
  \texttt{pardo@northwestern.edu} \\
}

\begin{document}

\maketitle

\begin{abstract}
  While research in AI methods for music generation and analysis has grown in scope and impact, AI researchers' engagement with the ethical consequences of this work has not kept pace. To encourage such engagement, many publication venues have introduced optional or required ethics statements for AI research papers. Though some authors use these ethics statements to critically engage with the broader implications of their research, we find that the majority of ethics statements in the AI music literature do not appear to be effectively utilized for this purpose. In this work, we conduct a review of ethics statements across ISMIR, NIME, and selected prominent works in AI music from the past five years. 
  We then offer suggestions for both audio conferences and researchers for engaging with ethics statements in ways that foster meaningful reflection rather than formulaic compliance. 
\end{abstract}

\section{Introduction}\label{sec:intro}

AI music and audio models are gaining more traction and prominence in society every day. Conferences like the International Society for Music Information Retrieval (ISMIR) and the International Conference on Acoustics, Speech, and Signal Processing (ICASSP) have cited record high submissions over the past couple years. 
Text-to-music commercial generative models like Suno and Udio report tens of millions of users \cite{suno_case_2024}. Big Tech firms like Google \cite{agostinelli2023musiclm, magenta_rt} and Meta \cite{copet2023simple} continue to release their own music models. What does \textit{not} continue to grow at the same rate is an internal reflection on ethical issues of these models from the creators of the works. 

In recent years, computer science researchers have sought to address, engage with, and mitigate the negative impacts of their work through a change to the peer review process \cite{hecht2021s}. NeurIPS even implemented a mandatory broader impact statement in 2020 \cite{nanayakkara2021unpacking}, but it is now defunct and instead baked into a smaller aspect of a required checklist. Conferences such as ISMIR have recently begun implementing an optional ethics statement that does not count towards the word limit, citing ``the goal of encouraging more discussion of ethical considerations in the ISMIR research community.''\footnote{https://ismir2024.ismir.net/authors/call-for-papers} This is in line with a larger trend of conferences encouraging ethical reflection by allowing or requiring an extra page for ethics and broader impact statements, e.g., the International Conference on Machine Learning (ICML) and the ACM Conference on Fairness, Accountability, and Transparency (FAccT).
\footnote{https://icml.cc/Conferences/2025/CallForPapers, https://facctconference.org/2025/aguide 
}

These statements are a valuable opportunity for authors and creators of these models to engage with the ethical implications of their work. However, when optional, only a minority of people choose to use them: for instance, only 28\% of papers published at ISMIR in 2024 used the page, and often they are not used for the designated purpose. This is reflective of a wider issue: a recent review of generative audio works found that less than 10\% of papers discuss any potential negative impacts of their work \cite{barnett2023ethical}. This is not indicative of a lack of potential harm---the work also details a suite of harms including climate impact, copyright infringement, deepfakes, and creativity stifling---but rather a failure by the research community to ethically engage with their own work. 

We see this moment as an opportunity: as ethics statements are increasingly incorporated into the research process, 
there is space to support researchers in approaching them productively. We seek to contribute to that effort by offering guidance, examples, and suggestions for engaging with ethics statements in ways that foster meaningful reflection rather than formulaic compliance. In this work, we analyze how authors in music-AI research are currently using these ethics statements, outline key categories of potential harm that may warrant discussion, and provide calls to action to both conferences and researchers. The goal of this work is not to encourage formulaic statements, but to support authors in developing thoughtful, substantive accounts of ethical implications of their work in order to become better scientists.




\section{Current State of Ethics in AI Music: A Mix of the Good, Bad, and Ugly}

\subsection{ISMIR 2024: An Optional Bonus Page} 

\begin{table}[h]
\centering
\begin{tabular}{lcl}
\toprule
\multicolumn{3}{c}{\textbf{Harms Discussed in ISMIR Ethics Statements}}\\
\midrule
\textbf{Harm} & \textbf{Count} & \textbf{Citation}\\
\midrule
Copyright Infringement & 13 (45\%) & \cite{barnett2024exploring, batlle2024towards, bukey2024just, gao2024variation, lenz2024cadenza, malandro2024composer, nguyen2024exploring, preniqi2024automatic, ramoneda2024music, riley2024gaps, sowula2024mosaikbox, suda2024fruitsmusic, wu2024melodyt5}\\
Labor Displacement & 8 (28\%) & \cite{lenz2024cadenza, nercessian2024generating, ramoneda2024towards, ramoneda2024music, riley2024gaps, sowula2024mosaikbox, tal2024joint, vanka2024diff}\\
General Bias & 7 (24\%) & \cite{batlle2024towards, comunita2024specmaskgit, evans2024long, lenz2024cadenza, maia2024investigating, preniqi2024automatic, sowula2024mosaikbox, weck2024muchomusic}\\
Cultural Appropriation & 7 (24\%) & \cite{barnett2024exploring, batlle2024towards, evans2024long, lenz2024cadenza, shankar2024saraga, weck2024muchomusic}\\
Voice Replication/Impersonation & 5 (17\%) & \cite{bukey2024just, malandro2024composer, nercessian2024generating, shikarpur2024hierarchical, suda2024fruitsmusic}\\
Western Bias & 5 (17\%) &  \cite{affolter2024utilizing, beyer2024end, bukey2024just, tal2024joint, watcharasupat2024stem}\\
Climate Impact & 3 (10\%) & \cite{batlle2024towards, holzapfel2024green, lenz2024cadenza}\\
Data Scraping & 3 (10\%) & \cite{batlle2024towards, flexer2024validity, tal2024joint}\\
Privacy Concerns & 3 (10\%) & \cite{maia2024investigating, suda2024fruitsmusic, wu2024melodyt5}\\
Sustainability & 2 (7\%) & \cite{batlle2024towards, nercessian2024generating}\\
Authorship & 2 (7\%) & \cite{barnett2024exploring, batlle2024towards}\\
Other & 6 (21\%)& \cite{barnett2024exploring, desblancs2024real, evans2024long, ramoneda2024towards, ramoneda2024music, vanka2024diff}\\
\bottomrule
\end{tabular}
\caption{Harms discussed in the 29 effectively utilized ISMIR ethics statements, organized by count of use, percent (of 29), and citations of papers discussing those harms.}
\label{tab:ismir_harms}
\end{table}

The optional extra page ISMIR provides which we describe above (Sec. \ref{sec:intro}) gives researchers an option to engage with the broader impact and ethical implications of their work without having to cut any content from the main body of the paper. Even with this added bonus page option, only $n=37/133 (28\%)$ of works wrote one. Further, the average statement was 1.8 paragraphs long, and only 169 words long (median of 148)---relative to the larger paper, these are very short. Of these $37$ ethics statements, $6$ ($16\%$) limited their ethical discussion to a statement that their project was approved by IRB. Two ($n=2; 5\%$) other works used this space not to engage ethically with their work, but to state they did not see any potential for ethical issues 
or to describe the history and ownership of the data they were using
, 
without further elaboration of the ethical impact of their own work. That leaves $29/133 (22\%)$ papers published in ISMIR proceedings that detailed potential ethical implications of their work using this bonus page.

Across these 29 ethics statements, there were 17 unique harms discussed (detailed in Table \ref{tab:ismir_harms}). The most prominent was the potential for copyright infringement ($n=13; 45\%$) through use of released datasets or models trained on those datasets. The second most discussed was the potential for labor displacement ($n=8; 28\%$) as a result of these tools being used in industry in place of artists or other music and sound-related professions such as sound engineers. Tied for third ($n=7; 24\%$) were general unspecified bias of the training data and the potential for misappropriation of cultures through unknowingly using these tools. 5 $(17\%)$ papers each discussed the potential harm for voice replication or impersonation through misuse of these models, as well as a Western bias of music, sounds, and voices in the training datasets. Only three papers ($10\%$) mentioned the climate impact, which is likely relevant for the majority of these works. Similarly, only three works each acknowledged data scraping and privacy as ethical issues. All other harms were mentioned only once or twice and include sustainability of business models, questions of authorship in AI creations, stifling of creativity, homogenization of music, memorization of training data, loss of agency, replicability of scientific works, and the potential for false positives in a deepfake detection algorithm. 

\subsection{NIME 2024/2025: A Mandatory Statement in the Main Text} %

The International Conference on New Interfaces for Musical Expression (NIME)\footnote{https://nime.org/} collects research related to musical interface design as well as artistic performances using new musical interfaces. The conference maintains a set of principles and a code of practice for maintaining ethical research \cite{morreale2023nime}. It also includes a strict requirement for ethics statements \footnote{https://github.com/NIME-conference/nime-template/releases/tag/v2024.12.02}, using the following text to describe guidelines for what authors should include:

\begin{quote}
    ``...information regarding sources of funding, potential conflicts of interest (financial or non-financial), informed consent if the research involved human participants, statement on welfare of animals if the research involved animals or any other information or context that helps ethically situate your research.''
\end{quote}

Of the 50 papers in the 2024 and 2025 NIME proceedings containing the phrases ``artificial intelligence,'' ``machine learning,'' or ``neural networks,'' 49 papers included the required ethics statements. However, without a dedicated page, many of the ethics statements stand as a single paragraph conforming to the exact wording of the guidelines, with few delving deeper than funding and recruitment sources. Themes among these ethics statements include the use of the author's own music as data and the use of a system as a tool specifically designed for the author. More generally, these ethics sections include conflicts of interest, participant informed consent, and ethical data sourcing (e.g., creating one’s own datasets or using appropriately licensed material). More detailed reflections tend to arise in projects involving tangible hardware, where sustainability and materials are emphasized, or in practice-based collaborations with artists, where authorship and participation are foregrounded.

Exceptions to this pattern occur in NIME papers related to the philosophical aspects of sustainability and practice in musical interfaces. For example, an analysis of the challenges of preserving deep generative models for NIMEs uses its Ethical Standards section to articulate the paper's goals of minimizing energy usage when recreating models, as well as to consider the authors' inclusion of the original authors of the systems compared \cite{clarke2025longevity}. 

It is important to note that NIME proceedings are diverse, spanning augmented instruments, installations, neural audio plugins, HCI analyses, and human–AI co-creation systems to name a few. Compared to model- and system-driven venues like ICASSP or ISMIR, NIME papers often center specific users (e.g., musicians, performers), practice-based methods, and artistic reflections. As a result, ethical considerations are frequently discussed outside of the ethics section, either in separately dedicated sections of the paper (\cite{nime2024_42, nime2024_81, nime2024_49}) or in the discussion/conclusion sections (\cite{nime2024_62}). 

\subsection{Prominent Music AI Works of the 2020s: A Mix of Everything}

\begin{table}[h]
\centering
\begin{tabular}{lcl}
\toprule
\multicolumn{3}{c}{\textbf{Prominent Music AI Works of the 2020s Surveyed in this Work}}\\
\midrule
\textbf{Harm/Topic Discussed in Ethics Statement} & \textbf{Count} & \textbf{Citation}\\
\midrule
Labor/Economic Impact & 7 (54\%) & \cite{copet2023simple, donahue2023singsong, novack2024ditto, schneider2023mo, tal2024joint, thickstun2023anticipatory, wu2024music}\\
Legal Acquisition of Data & 6 (46\%) & \cite{copet2023simple, melechovsky2023mustango, nistal2024diff, novack2024ditto,tal2024joint,  wu2024music}\\
General Bias & 6 (46\%) & \cite{agostinelli2023musiclm, copet2023simple, evans2024fast, evans2024long, huang2023noise2music, tal2024joint}\\
Western Bias & 4 (31\%) & \cite{copet2023simple, huang2023noise2music, tal2024joint, thickstun2023anticipatory}\\
Loss of Agency/Creativity Stifling & 4 (31\%) & \cite{donahue2023singsong, novack2024ditto, thickstun2023anticipatory, wu2024music}\\
Voice Replication/Impersonation & 2 (15\%) & \cite{donahue2023singsong, wu2024music}\\
Copyright & 2 (15\%) &  \cite{schneider2023mo, thickstun2023anticipatory}\\
Cultural Appropriation & 2 (13\%) & \cite{agostinelli2023musiclm, huang2023noise2music}\\
\midrule
\multirow{2}{*}{No Ethics Statement} & \multirow{2}{*}{12} & \cite{caillon2021rave, chen2024musicldm, dhariwal2020jukebox, evans2025stable, garcia2023vampnet, lam2023efficient}\\
& & \cite{ lan2024high, ning2025diffrhythm, novack2024ditto2, novack2025fast,  nistal2024improving, parker2024stemgen}\\
\bottomrule
\end{tabular}
\caption{Harms (or topics) discussed in the 13 ethics/broader impact sections of the 25 prominent Music AI papers surveyed in this work, organized by count of use, percent (of 13), and citations of papers discussing those harms.}
\label{tab:prominent_paper_harms}
\end{table}

In addition to these two music conferences, we also surveyed prominent music generation papers from the last five years; these papers were mainly produced by established academic research groups, Big Tech, or collaborations between the two. We used as a starting point the 16 music generative modeling papers recently assessed for openness by MusGO \cite{batlle2025musgo}, which cover ``a diverse set of music-generative
models'' and ``a variety of architectures commonly used within the field.'' We bolstered this set with 9 additional papers identified by our team as high-impact or otherwise representative of state-of-the-art methods, none of which overlapped with 2024 ISMIR or NIME. The final set can be found in Table \ref{tab:prominent_paper_harms}. Of these 25 works, $n=13; (52\%)$ had some form of ethics statement. These were typically a short paragraph or section after the conclusion $(n=10;77\%)$. Additionally, $n=2$ were written as part of the discussion, and $n=1$ was written in the appendix. On average these were slightly longer than ISMIR statements---1.9 paragraphs and 281 words---though this was driven by one much longer and more effective ethics statement \cite{thickstun2023anticipatory} which was 11 paragraphs and 1,576 words; the median was on par with ISMIR at 147 words. 

Many of these ethics statements reflect a trend in which researchers seek to preemptively defend their work from criticism rather than to acknowledge and reckon with potential harms. None of these 25 papers mentioned the potential impact on the environment. 
Only two works ($n=2$) in this corpus discussed the potential for copyright infringement \cite{schneider2023mo, thickstun2023anticipatory}, which is a stark contrast to the $45\%$ of ISMIR ethics statements that discussed this. It is potentially indicative of industry wariness to even acknowledge the potential for copyright infringement, seeing as $n=20/25 (75\%)$ of these works had at least one industry author and $n=16/25 (64\%)$ had a majority of industry affiliated authors. Thickstun et al. \cite{thickstun2023anticipatory} assert ``we are certain that generative models of music are similarly capable of plagiarism,'' citing works on memorization in text \cite{carlini2022quantifying} and vision \cite{somepalli2023diffusion}, which has now been echoed by a works in music \cite{barnett2024exploring, batlle2024towards}. Works trained on massive amounts of data such as VampNet \cite{garcia2023vampnet} and Mo\^{u}sai \cite{schneider2023mo} notably lack this acknowledgment in their papers. VampNet used hundreds of thousands of scraped songs for training data and had no ethics statement, and Mo\^{u}sai scraped roughly 50,000 songs from Spotify---though the authors disclose their data acquisition process and open-source their scraping code. While this transparency is laudable, the authors do not use the ethics statement to engage with the implications of their use of copyrighted data, but rather to defend this use as technically legal or ``exempted from copyright infringement.''

About half of these statements focused on stating in explicit terms that they acquired their data through legal means ($n=6; 46\%$), sometimes without engaging critically beyond stating that ($n=2$). Similarly, two works ($n=2$) only briefly mentioned that their training data has biases without engaging beyond a vague description of what that means. The most commonly discussed harm was labor impact on artists, with $n=7;54\%$ of statements discussing this \cite{copet2023simple, donahue2023singsong, novack2024ditto, schneider2023mo, tal2024joint, thickstun2023anticipatory, wu2024music}. As \cite{tal2024joint} notes, ``generative models could potentially create an unbalanced competitive environment for artists, a problem that is yet to be solved;'' this is a harm that could be mentioned in each and every one of the 25 papers surveyed in this section. 

Four of these works \cite{copet2023simple, huang2023noise2music, tal2024joint, thickstun2023anticipatory} discussed the potential harm resulting from training on a predominance of Western music data, which Thickstun et al. note ``inevitably biases the model’s predictions towards infilling completions that follow Western rules of composition'' \cite{thickstun2023anticipatory}. This is an example of more critically engaging with the biases present in your dataset and discussing potential implications of that, rather than just vaguely gesturing towards the idea of bias.




\section{Call to Action: Changes to this Process}

This review of the recent use of ethics statements has made two key aspects clear: (1) \textbf{ethics statements are underutilized} both in terms of writing them at all and discussing a comprehensive suite of relevant harms, and (2) researchers are frequently \textbf{using these statements ineffectively} by not engaging with the ethical implications of their work. We believe there are a series of changes that could right both of these wrongs, but they require collective action by the research community across various stages of the research process. 

\textbf{\textit{Audio conferences}} should start requiring an extra page allotted for ethics statements instead of including them as an optional page (or even not allowing space for them at all). However, 
these statements should not be a regurgitation of the ethical guidelines provided by the conference. To help avoid this, the call for papers pages should provide excellent examples of ethics statements from previous years (such as Thickstun et al. \cite{thickstun2023anticipatory}, Batlle-Roca et al. \cite{batlle2024towards}, or Barnett et al. \cite{barnett2024exploring}), so as to call attention to ethically conducted work and authors who properly engage critically with their work. The ethics statements could also be required as part of the abstract registration a week before the conference in order to encourage authors to write these statements thoughtfully and not just as an afterthought. NeurIPS especially should reconsider its Broader Impact Criteria rather than baking it into a long and tedious checklist that does not reach the readers of these papers. Checklists like this prompt researchers to believe ethics statements are about bureaucracy rather than ethics. 

\textbf{\textit{Researchers}} in this field should start viewing these statements as an opportunity to begin meaningfully engaging with the broader implications of their own work. Though they typically come at the end of the work in reflection, one should begin thinking about the ethical implications of the work at the start of the research process. 
In addition to detailing potential harms in ethics statements, one can describe steps taken to actively mitigate potential harms or even harms the work may be actively addressing. If it is unclear where to start, they should review a set of harms covered by other works in the field (such as the ones described in Tables \ref{tab:ismir_harms}-\ref{tab:prominent_paper_harms} or in a review of harms of these models \cite{barnett2023ethical}). They should identify the harms that could be applicable and detail ways in which the work is vulnerable to those harms, may perpetuate those harms, or has even taken action to fight against these harms. Authors should not merely list these off as a checklist, but take the opportunity to engage with them. For instance, if the researcher performed any training runs on GPUs they can estimate the environmental cost of a training run: number of GPUs $\times$ energy output of GPUs (W) $\times$ hours of training time to get an energy cost in kilowatt-hours (kWh); then the author can disclose this for transparency and to encourage their peers to do so as well \cite{holzapfel2024green}.

\section{Conclusion and Future Work}

In this work we examined the usage of ethics statements in music AI works across ISMIR, NIME, and other prominent works in the domain. We found that though some authors are using these effectively to critically engage with the broader implications of their work, there is a dearth of usage and even more so proper usage of these statements. We identify harms that have been discussed which are widely applicable, and we present suggestions to improve this process at large. We recognize we can be a positive force in this domain as researchers intimately familiar with both the ethical implications of these works and the domain itself. In the future we plan to release a more interactive, open, and living document guide to assist authors with engaging with these harms. We envision doing so through a decision-tree type mechanism (e.g., did you use training data in your work? If so these are some potential harms.) If you would like to contribute, please reach out to the authors of this work.

\begin{ack}
This work was supported in part by USA National Science Foundation award numbers 2222369 and 2300633.
\end{ack}

\bibliography{references}

\begin{thebibliography}{10}

\bibitem{affolter2024utilizing}
Joanne Affolter and Martin~A Rohrmeier.
\newblock Utilizing listener-provided tags for music emotion recognition: A data-driven approach.
\newblock In {\em Proceedings of the 25th International Society for Music Information Retrieval Conference (ISMIR)}, 2024.

\bibitem{agostinelli2023musiclm}
Andrea Agostinelli, Timo~I Denk, Zal{\'a}n Borsos, Jesse Engel, Mauro Verzetti, Antoine Caillon, Qingqing Huang, Aren Jansen, Adam Roberts, Marco Tagliasacchi, et~al.
\newblock Musiclm: Generating music from text.
\newblock {\em arXiv preprint arXiv:2301.11325}, 2023.

\bibitem{nime2024_42}
Jack Armitage, Victor Shepardson, and Thor Magnusson.
\newblock Tölvera: Composing with basal agencies.
\newblock pages 282--291, September 2024.

\bibitem{barnett2023ethical}
Julia Barnett.
\newblock The ethical implications of generative audio models: A systematic literature review.
\newblock In {\em Proceedings of the 2023 AAAI/ACM Conference on AI, Ethics, and Society}, pages 146--161, 2023.

\bibitem{barnett2024exploring}
Julia Barnett, Hugo~Flores Garcia, and Bryan Pardo.
\newblock Exploring musical roots: Applying audio embeddings to empower influence attribution for a generative music model.
\newblock In {\em Proceedings of the 25th International Society for Music Information Retrieval Conference (ISMIR)}, 2024.

\bibitem{batlle2025musgo}
Roser Batlle-Roca, Laura Ib{\'a}{\~n}ez-Mart{\'\i}nez, Xavier Serra, Emilia G{\'o}mez, and Mart{\'\i}n Rocamora.
\newblock Musgo: A community-driven framework for assessing openness in music-generative ai.
\newblock {\em arXiv preprint arXiv:2507.03599}, 2025.

\bibitem{batlle2024towards}
Roser Batlle-Roca, Wei-Hsiang Liao, Xavier Serra, Yuki Mitsufuji, and Emilia G{\'o}mez.
\newblock Towards assessing data replication in music generation with music similarity metrics on raw audio.
\newblock In {\em Proceedings of the 25th International Society for Music Information Retrieval Conference (ISMIR)}, 2024.

\bibitem{beyer2024end}
Tim Beyer and Angela Dai.
\newblock End-to-end piano performance-midi to score conversion with transformers.
\newblock In {\em Proceedings of the 25th International Society for Music Information Retrieval Conference (ISMIR)}, 2024.

\bibitem{bukey2024just}
Irmak Bukey, Michael Feffer, and Chris Donahue.
\newblock Just label the repeats for in-the-wild audio-to-score alignment.
\newblock In {\em Proceedings of the 25th International Society for Music Information Retrieval Conference (ISMIR)}, 2024.

\bibitem{caillon2021rave}
Antoine Caillon and Philippe Esling.
\newblock Rave: A variational autoencoder for fast and high-quality neural audio synthesis.
\newblock {\em arXiv preprint arXiv:2111.05011}, 2021.

\bibitem{carlini2022quantifying}
Nicholas Carlini, Daphne Ippolito, Matthew Jagielski, Katherine Lee, Florian Tramer, and Chiyuan Zhang.
\newblock Quantifying memorization across neural language models.
\newblock In {\em The Eleventh International Conference on Learning Representations}, 2022.

\bibitem{chen2024musicldm}
Ke~Chen, Yusong Wu, Haohe Liu, Marianna Nezhurina, Taylor Berg-Kirkpatrick, and Shlomo Dubnov.
\newblock Musicldm: Enhancing novelty in text-to-music generation using beat-synchronous mixup strategies.
\newblock In {\em ICASSP 2024-2024 IEEE International Conference on Acoustics, Speech and Signal Processing (ICASSP)}, pages 1206--1210. IEEE, 2024.

\bibitem{clarke2025longevity}
Isaac Clarke, Francesco~Ardan Dal~R{\'\i}, and Raul Masu.
\newblock Longevity of deep generative models in nime: Challenges and practices for reactivation.
\newblock In {\em Proceedings of the International Conference on New Interfaces for Musical Expression}, pages 224--230, 2025.

\bibitem{comunita2024specmaskgit}
Marco Comunit{\`a}, Zhi Zhong, Akira Takahashi, Shiqi Yang, Mengjie Zhao, Koichi Saito, Yukara Ikemiya, Takashi Shibuya, Shusuke Takahashi, and Yuki Mitsufuji.
\newblock Specmaskgit: Masked generative modeling of audio spectrograms for efficient audio synthesis and beyond.
\newblock In {\em Proceedings of the 25th International Society for Music Information Retrieval Conference (ISMIR)}, 2024.

\bibitem{copet2023simple}
Jade Copet, Felix Kreuk, Itai Gat, Tal Remez, David Kant, Gabriel Synnaeve, Yossi Adi, and Alexandre D{\'e}fossez.
\newblock Simple and controllable music generation.
\newblock {\em Advances in Neural Information Processing Systems}, 36:47704--47720, 2023.

\bibitem{desblancs2024real}
Dorian Desblancs, Gabriel Meseguer-Brocal, Romain Hennequin, and Manuel Moussallam.
\newblock From real to cloned singer identification.
\newblock In {\em Proceedings of the 25th International Society for Music Information Retrieval Conference (ISMIR)}, 2024.

\bibitem{dhariwal2020jukebox}
Prafulla Dhariwal, Heewoo Jun, Christine Payne, Jong~Wook Kim, Alec Radford, and Ilya Sutskever.
\newblock Jukebox: A generative model for music.
\newblock {\em arXiv preprint arXiv:2005.00341}, 2020.

\bibitem{donahue2023singsong}
Chris Donahue, Antoine Caillon, Adam Roberts, Ethan Manilow, Philippe Esling, Andrea Agostinelli, Mauro Verzetti, Ian Simon, Olivier Pietquin, Neil Zeghidour, et~al.
\newblock Singsong: Generating musical accompaniments from singing.
\newblock {\em arXiv preprint arXiv:2301.12662}, 2023.

\bibitem{evans2024fast}
Zach Evans, CJ~Carr, Josiah Taylor, Scott~H Hawley, and Jordi Pons.
\newblock Fast timing-conditioned latent audio diffusion.
\newblock In {\em Forty-first International Conference on Machine Learning}, 2024.

\bibitem{evans2024long}
Zach Evans, Julian~D Parker, CJ~Carr, Zack Zukowski, Josiah Taylor, and Jordi Pons.
\newblock Long-form music generation with latent diffusion.
\newblock In {\em Proceedings of the 25th International Society for Music Information Retrieval Conference (ISMIR)}, 2024.

\bibitem{evans2025stable}
Zach Evans, Julian~D Parker, CJ~Carr, Zack Zukowski, Josiah Taylor, and Jordi Pons.
\newblock Stable audio open.
\newblock In {\em ICASSP 2025-2025 IEEE International Conference on Acoustics, Speech and Signal Processing (ICASSP)}, pages 1--5. IEEE, 2025.

\bibitem{flexer2024validity}
Arthur Flexer.
\newblock On the validity of employing chatgpt for distant reading of music similarity.
\newblock In {\em Proceedings of the 25th International Society for Music Information Retrieval Conference (ISMIR)}, 2024.

\bibitem{gao2024variation}
Chenyu Gao, Federico Reuben, and Tom Collins.
\newblock Variation transformer: New datasets, models, and comparative evaluation for symbolic music variation generation.
\newblock In {\em Proceedings of the 25th International Society for Music Information Retrieval Conference (ISMIR)}, 2024.

\bibitem{garcia2023vampnet}
Hugo~Flores Garcia, Prem Seetharaman, Rithesh Kumar, and Bryan Pardo.
\newblock Vampnet: Music generation via masked acoustic token modeling.
\newblock {\em arXiv preprint arXiv:2307.04686}, 2023.

\bibitem{hecht2021s}
Brent Hecht, Lauren Wilcox, Jeffrey~P Bigham, Johannes Sch{\"o}ning, Ehsan Hoque, Jason Ernst, Yonatan Bisk, Luigi De~Russis, Lana Yarosh, Bushra Anjum, et~al.
\newblock It's time to do something: Mitigating the negative impacts of computing through a change to the peer review process.
\newblock {\em arXiv preprint arXiv:2112.09544}, 2021.

\bibitem{holzapfel2024green}
Andre Holzapfel, Anna-Kaisa Kaila, and Petra J{\"a}{\"a}skel{\"a}inen.
\newblock Green mir?: Investigating computational cost of recent music-ai research in ismir.
\newblock In {\em Proceedings of the 25th International Society for Music Information Retrieval Conference (ISMIR)}, 2024.

\bibitem{huang2023noise2music}
Qingqing Huang, Daniel~S Park, Tao Wang, Timo~I Denk, Andy Ly, Nanxin Chen, Zhengdong Zhang, Zhishuai Zhang, Jiahui Yu, Christian Frank, et~al.
\newblock Noise2music: Text-conditioned music generation with diffusion models.
\newblock {\em arXiv preprint arXiv:2302.03917}, 2023.

\bibitem{lam2023efficient}
Max~WY Lam, Qiao Tian, Tang Li, Zongyu Yin, Siyuan Feng, Ming Tu, Yuliang Ji, Rui Xia, Mingbo Ma, Xuchen Song, et~al.
\newblock Efficient neural music generation.
\newblock {\em Advances in Neural Information Processing Systems}, 36:17450--17463, 2023.

\bibitem{lan2024high}
Gael~Le Lan, Bowen Shi, Zhaoheng Ni, Sidd Srinivasan, Anurag Kumar, Brian Ellis, David Kant, Varun Nagaraja, Ernie Chang, Wei-Ning Hsu, et~al.
\newblock High fidelity text-guided music editing via single-stage flow matching.
\newblock {\em arXiv preprint arXiv:2407.03648}, 2024.

\bibitem{lenz2024cadenza}
Julian Lenz and Anirudh Mani.
\newblock Cadenza: A generative framework for expressive musical ideas and variations.
\newblock In {\em Proceedings of the 25th International Society for Music Information Retrieval Conference (ISMIR)}, 2024.

\bibitem{nime2024_81}
Yikai Li and Ge~Wang.
\newblock Chai => interactive ai tools in chuck.
\newblock pages 553--559, September 2024.

\bibitem{maia2024investigating}
Lucas~S Maia, Richa Namballa, Mart{\'\i}n Rocamora, Magdalena Fuentes, and Carlos Guedes.
\newblock Investigating time-line-based music traditions with field recordings: a case study of candombl{\'e} bell patterns.
\newblock In {\em Proceedings of the 25th International Society for Music Information Retrieval Conference (ISMIR)}, 2024.

\bibitem{malandro2024composer}
Martin~E Malandro.
\newblock Composer's assistant 2: Interactive multi-track midi infilling with fine-grained user control.
\newblock In {\em Proceedings of the 25th International Society for Music Information Retrieval Conference (ISMIR)}, 2024.

\bibitem{melechovsky2023mustango}
Jan Melechovsky, Zixun Guo, Deepanway Ghosal, Navonil Majumder, Dorien Herremans, and Soujanya Poria.
\newblock Mustango: Toward controllable text-to-music generation.
\newblock {\em arXiv preprint arXiv:2311.08355}, 2023.

\bibitem{morreale2023nime}
Fabio Morreale, Nicolas Gold, C{\'e}cile Chevalier, and Raul Masu.
\newblock Nime principles \& code of practice on ethical research.
\newblock 2023.

\bibitem{nanayakkara2021unpacking}
Priyanka Nanayakkara, Jessica Hullman, and Nicholas Diakopoulos.
\newblock Unpacking the expressed consequences of ai research in broader impact statements.
\newblock In {\em Proceedings of the 2021 AAAI/ACM Conference on AI, Ethics, and Society}, pages 795--806, 2021.

\bibitem{nercessian2024generating}
Shahan Nercessian, Johannes Imort, Ninon Devis, and Frederik Blang.
\newblock Generating sample-based musical instruments using neural audio codec language models.
\newblock In {\em Proceedings of the 25th International Society for Music Information Retrieval Conference (ISMIR)}, 2024.

\bibitem{nguyen2024exploring}
Ngan~VT Nguyen, Elizabeth~AM Acosta, Tommy Dang, and David~RW Sears.
\newblock Exploring internet radio across the globe with the mirage online dashboard.
\newblock In {\em Proceedings of the 25th International Society for Music Information Retrieval Conference (ISMIR)}, 2024.

\bibitem{ning2025diffrhythm}
Ziqian Ning, Huakang Chen, Yuepeng Jiang, Chunbo Hao, Guobin Ma, Shuai Wang, Jixun Yao, and Lei Xie.
\newblock Diffrhythm: Blazingly fast and embarrassingly simple end-to-end full-length song generation with latent diffusion.
\newblock {\em arXiv preprint arXiv:2503.01183}, 2025.

\bibitem{nistal2024diff}
Javier Nistal, Marco Pasini, Cyran Aouameur, Maarten Grachten, and Stefan Lattner.
\newblock Diff-a-riff: Musical accompaniment co-creation via latent diffusion models.
\newblock In {\em Proceedings of the 25th International Society for Music Information Retrieval Conference (ISMIR)}, 2024.

\bibitem{nistal2024improving}
Javier Nistal, Marco Pasini, and Stefan Lattner.
\newblock Improving musical accompaniment co-creation via diffusion transformers.
\newblock {\em arXiv preprint arXiv:2410.23005}, 2024.

\bibitem{novack2025fast}
Zachary Novack, Zach Evans, Zack Zukowski, Josiah Taylor, CJ~Carr, Julian Parker, Adnan Al-Sinan, Gian~Marco Iodice, Julian McAuley, Taylor Berg-Kirkpatrick, et~al.
\newblock Fast text-to-audio generation with adversarial post-training.
\newblock {\em arXiv preprint arXiv:2505.08175}, 2025.

\bibitem{novack2024ditto2}
Zachary Novack, Julian McAuley, Taylor Berg-Kirkpatrick, and Nicholas Bryan.
\newblock Ditto-2: Distilled diffusion inference-time t-optimization for music generation.
\newblock {\em arXiv preprint arXiv:2405.20289}, 2024.

\bibitem{novack2024ditto}
Zachary Novack, Julian McAuley, Taylor Berg-Kirkpatrick, and Nicholas~J Bryan.
\newblock Ditto: Diffusion inference-time t-optimization for music generation.
\newblock {\em arXiv preprint arXiv:2401.12179}, 2024.

\bibitem{parker2024stemgen}
Julian~D Parker, Janne Spijkervet, Katerina Kosta, Furkan Yesiler, Boris Kuznetsov, Ju-Chiang Wang, Matt Avent, Jitong Chen, and Duc Le.
\newblock Stemgen: A music generation model that listens.
\newblock In {\em ICASSP 2024-2024 IEEE International Conference on Acoustics, Speech and Signal Processing (ICASSP)}, pages 1116--1120. IEEE, 2024.

\bibitem{preniqi2024automatic}
Vjosa Preniqi, Iacopo Ghinassi, Julia Ive, Kyriaki Kalimeri, and Charalampos Saitis.
\newblock Automatic detection of moral values in music lyrics.
\newblock In {\em Proceedings of the 25th International Society for Music Information Retrieval Conference (ISMIR)}, 2024.

\bibitem{nime2024_62}
Nicola Privato, Victor Shepardson, Giacomo Lepri, and Thor Magnusson.
\newblock Stacco: Exploring the embodied perception of latent representations in neural synthesis.
\newblock pages 424--431, September 2024.

\bibitem{ramoneda2024towards}
Pedro Ramoneda, Vsevolod Eremenko, Alexandre D'Hooge, Emilia Parada-Cabaleiro, and Xavier Serra.
\newblock Towards explainable and interpretable musical difficulty estimation: a parameter-efficient approach.
\newblock In {\em Proceedings of the 25th International Society for Music Information Retrieval Conference (ISMIR)}, 2024.

\bibitem{ramoneda2024music}
Pedro Ramoneda, Martin Rocamora, and Taketo Akama.
\newblock Music proofreading with refinpaint: where and how to modify compositions given context.
\newblock In {\em Proceedings of the 25th International Society for Music Information Retrieval Conference (ISMIR)}, 2024.

\bibitem{riley2024gaps}
Xavier Riley, Zixun Guo, Drew Edwards, and Simon Dixon.
\newblock Gaps: A large and diverse classical guitar dataset and benchmark transcription model.
\newblock In {\em Proceedings of the 25th International Society for Music Information Retrieval Conference (ISMIR)}, 2024.

\bibitem{schneider2023mo}
Flavio Schneider, Ojasv Kamal, Zhijing Jin, and Bernhard Sch{\"o}lkopf.
\newblock Mo$\backslash$\^{} usai: Text-to-music generation with long-context latent diffusion.
\newblock {\em arXiv preprint arXiv:2301.11757}, 2023.

\bibitem{nime2024_49}
Nicholas Shaheed and Ge~Wang.
\newblock I am sitting in a (latent) room.
\newblock pages 333--338, September 2024.

\bibitem{shankar2024saraga}
Adithi Shankar, Gen{\'\i}s Plaja-Roglans, Thomas Nuttall, Mart{\'\i}n Rocamora, and Xavier Serra.
\newblock Saraga audiovisual: a large multimodal open data collection for the analysis of carnatic music.
\newblock In {\em Proceedings of the 25th International Society for Music Information Retrieval Conference (ISMIR)}, 2024.

\bibitem{shikarpur2024hierarchical}
Nithya Shikarpur, Krishna~Maneesha Dendukuri, Yusong Wu, Antoine Caillon, and Cheng-Zhi~Anna Huang.
\newblock Hierarchical generative modeling of melodic vocal contours in hindustani classical music.
\newblock In {\em Proceedings of the 25th International Society for Music Information Retrieval Conference (ISMIR)}, 2024.

\bibitem{somepalli2023diffusion}
Gowthami Somepalli, Vasu Singla, Micah Goldblum, Jonas Geiping, and Tom Goldstein.
\newblock Diffusion art or digital forgery? investigating data replication in diffusion models.
\newblock In {\em Proceedings of the IEEE/CVF conference on computer vision and pattern recognition}, pages 6048--6058, 2023.

\bibitem{sowula2024mosaikbox}
Robert Sowula and Peter Knees.
\newblock Mosaikbox: Improving fully automatic dj mixing through rule-based stem modification and precise beat-grid estimation.
\newblock In {\em Proceedings of the 25th International Society for Music Information Retrieval Conference (ISMIR)}, 2024.

\bibitem{suda2024fruitsmusic}
Hitoshi Suda, Shunsuke Yoshida, Tomohiko Nakamura, Satoru Fukayama, and Jun Ogata.
\newblock Fruitsmusic: A real-world corpus of japanese idol-group songs.
\newblock In {\em Proceedings of the 25th International Society for Music Information Retrieval Conference (ISMIR)}, 2024.

\bibitem{tal2024joint}
Or~Tal, Alon Ziv, Itai Gat, Felix Kreuk, and Yossi Adi.
\newblock Joint audio and symbolic conditioning for temporally controlled text-to-music generation.
\newblock In {\em Proceedings of the 25th International Society for Music Information Retrieval Conference (ISMIR)}, 2024.

\bibitem{magenta_rt}
Lyria Team.
\newblock Magenta realtime.
\newblock 2025.

\bibitem{thickstun2023anticipatory}
John Thickstun, David Hall, Chris Donahue, and Percy Liang.
\newblock Anticipatory music transformer.
\newblock {\em arXiv preprint arXiv:2306.08620}, 2023.

\bibitem{suno_case_2024}
{UMG Recordings, Inc. v. Suno, Inc. (2025); Main complaint filed 06/24/2024}.

\bibitem{vanka2024diff}
Soumya~Sai Vanka, Christian Steinmetz, Jean-Baptiste Rolland, Joshua Reiss, and George Fazekas.
\newblock Diff-mst: Differentiable mixing style transfer.
\newblock In {\em Proceedings of the 25th International Society for Music Information Retrieval Conference (ISMIR)}, 2024.

\bibitem{watcharasupat2024stem}
Karn~N Watcharasupat and Alexander Lerch.
\newblock A stem-agnostic single-decoder system for music source separation beyond four stems.
\newblock In {\em Proceedings of the 25th International Society for Music Information Retrieval Conference (ISMIR)}, 2024.

\bibitem{weck2024muchomusic}
Benno Weck, Ilaria Manco, Emmanouil Benetos, Elio Quinton, George Fazekas, and Dmitry Bogdanov.
\newblock Muchomusic: Evaluating music understanding in multimodal audio-language models.
\newblock In {\em Proceedings of the 25th International Society for Music Information Retrieval Conference (ISMIR)}, 2024.

\bibitem{wu2024melodyt5}
Shangda Wu, Yashan Wang, Xiaobing Li, Feng Yu, and Maosong Sun.
\newblock Melodyt5: A unified score-to-score transformer for symbolic music processing.
\newblock In {\em Proceedings of the 25th International Society for Music Information Retrieval Conference (ISMIR)}, 2024.

\bibitem{wu2024music}
Shih-Lun Wu, Chris Donahue, Shinji Watanabe, and Nicholas~J Bryan.
\newblock Music controlnet: Multiple time-varying controls for music generation.
\newblock {\em IEEE/ACM Transactions on Audio, Speech, and Language Processing}, 32:2692--2703, 2024.

\end{thebibliography}


\appendix

\section{Ethics Statement}

In this section we write our own ethics statement for this work, demonstrating both what we believe to be an effective ethics statement (\ref{sec:effective_ethics}), which engages with the specific potential harms related to the content of the work, and an ineffective ethics statement (\ref{sec:ineffective_ethics}), which briefly addresses ``boilerplate'' issues unrelated to the potential harms of this particular work.

\subsection{An Effective Ethics Statement}\label{sec:effective_ethics}

The primary concern we have with writing this work is that though we encourage ethical engagement with work, researchers may use our suggestions for harms to discuss in music AI papers as boxes to check in writing a formulaic ethics statement. As a result, though we would be encouraging more researchers to write ethics statements, we would be inadvertently contributing to the exact problem identified in this work which is ineffective usage of these statements. 

Another concern we have with this work is that our sample is biased in multiple ways. With regard to ISMIR and NIME, we focused only on the ethics statements and did not include the main body of text in the corpus which we analyzed---as a result, we may be categorizing works that critically engaged with the broader implications of their work in the main text as not doing so in the manner we focused on. The papers we selected to reflect a sample of 25 prominent music AI papers in the last five years was certainly sampled to those we believe to be ``prominent,'' and that is inherently biased to our own views. We attempted to circumvent this by basing the bulk of these works on those selected by \cite{batlle2025musgo} because we both agreed with the selection and it was then verified my two independent sets of authors, but this is a biased sample nonetheless. 

Finally we acknowledge that calling to attention instances where researchers did not write an ethics statement may make some authors of these works uneasy, and we are not attempting to shame anyone by doing so. We are focusing on works that are already prominent in the field in terms of citations or big names (e.g., Google and Meta), knowing this work will not have a deleterious impact on them so as to not unjustly critique any nascent works. By doing so we highlight that all works in music AI going forward should have ethics statements and these big names are no exception. Nevertheless we acknowledge that this could be viewed in a negative light by our peers, and reiterate causing discomfort was not our intent. 

We believe these outlined risks are far outweighed by the potential benefit we view of conducting this review.
 
\subsection{An Ineffective Ethics Statement}\label{sec:ineffective_ethics}

We obtained all of the papers analyzed in this work through legal means because they were published on conference proceedings or arXiv. There is likely bias in the sample we analyzed, and future work should seek to mitigate this bias. We may have reproduced copyrighted content in part by quoting these works, though it was likely not a legal infringement due to the research provision for fair use under US copyright law. 

This work did not have any human subjects so it was not necessary to obtain IRB approval.


\end{document}